\documentclass[fonts]{icst}
\usepackage{moreverb}
\usepackage{listings}
\usepackage{wrapfig}
\usepackage{color}
\usepackage{caption}

\usepackage{subcaption}
\usepackage[breaklinks,colorlinks,bookmarksopen,bookmarksnumbered,linkcolor=ICSTblue,citecolor=blue,urlcolor=ICSTblue]{hyperref}
\usepackage{breakurl}
\usepackage{doi}
\usepackage{makecell}
\usepackage[linesnumbered, algoruled, lined, boxed, ruled, algosection, norelsize]{algorithm2e}

\SetAlFnt{\small\sffamily}
\newcommand\BibTeX{{\rmfamily B\kern-.05em \textsc{i\kern-.025em b}\kern-.08em
T\kern-.1667em\lower.7ex\hbox{E}\kern-.125emX}}

\def\volumeyear{2019}

\journalname{XXXXXX}
\articletype{Research Article/Editorial}
 \def\volumeyear{XXXX}
 \def\volumenumber{XX}
  \def\issuemonth{XX-XX}
  \def\issuenumber{X}
  \def\articleid{XX}
  \def\copyrightholder{Author Name\textit{ et al.}, licensed to ICST}
  \copyrightnote{This is an open access article distributed under the terms of the Creative Commons Attribution license (\url{http://creativecommons.org/licenses/by/3.0/}), which permits unlimited use, distribution and reproduction in any medium so long as the original work is properly cited.}
  \def\articledoi{10.4108/XX.X.X.XX}
\received{XXXX}
  \accepted{XXXX}
  \published{XXXX}

\begin{document}

\runningheads{Alomari and Harbi}{Scalable Source Code Similarity Detection in Large Code Repositories}

\title{Scalable Source Code Similarity Detection in Large Code Repositories}

\author{Firas Alomari\affil{1}\fnoteref{1} Muhammed Harbi\affil{1}}

\address{\affilnum{1} Corporate Applications Department, Saudi Aramco, Dhahran, Saudi Arabia\\}

\abstract{Source code similarity are increasingly used in application development to identify clones, isolate bugs, and find copy-rights violations. Similar code fragments can be very problematic due to the fact that errors in the original code must be fixed in every copy. Other maintenance changes, such as extensions or patches, must be applied multiple times. Furthermore,  the diversity of coding styles and flexibility of modern languages makes it difficult and cost ineffective to manually inspect large code repositories. Therefore, detection is only feasible by automatic techniques. We present an efficient and scalable approach for similar code fragment identification based on source code control flow graphs fingerprinting. The source code is processed to generate control flow graphs that are then hashed to create a unique fingerprint of the code capturing semantics as well as syntax similarity. The fingerprints can then be efficiently stored and retrieved to perform similarity search between code fragments. Experimental results from our prototype implementation supports the validity of our approach and show its effectiveness and efficiency in comparison with other solutions.}


%\tnotetext[1]{Please ensure that you use the most up to date
%class file, available from EAI at \url{http://doc.eai.eu/publications/transactions/latex/}
%}

\fnotetext[1]{Corresponding author.  Email: \email{firas.alomari@aramco.com}}

\maketitle

Enterprise Resource Planning (ERP) systems are a fundamental part in most companies IT application portfolio. They provide a set of standardized software applications that handles interdisciplinary business processes across the entire value chain of an enterprise \cite{kremers2000enterprise,lee2003enterprise}. The potential of ERP systems to integrate business functions such as supply chain, financial accounting, or Human Resources has led to their widespread adoption. One such example is the SAP (Systems Applications and Products) ERP software which provides standard packages capturing "best business practices" \cite{themistocleous2001erp}. However, rapid and continuous changes in business requirements are forcing companies to continuously modify and enhance the standard functionality to meet their needs \cite{brehm2001tailoring}. Therefore, developers often need to modify a specific pieces of code from the standard ERP code base to satisfy their business scenario requirements. Specifically, SAP allows their customers to develop their own enhancements by using Advance Business Application Programming (ABAP) fourth-generation programming language \cite{keller2010official}. However, their best practices discourage against modifications to the standard code in the system and strictly control it through the "repairs" and enhancements concepts \cite{keller2003abap}. Consequently, driving developers to copy standard code and modify it to meet their requirements, thus, creating duplicate code or clones. 

Duplicate code or clones can have severe impacts on quality, re-usability and maintainability of a software system for several reasons \cite{juergens2009code,gupta2018survey}: \textbf{1)} Copy and paste operations are error prone and introduce defects, \textbf{2)} they increase the probability of bug propagation in new code, \textbf{3)} they lead to loss of design abstraction of the system decreasing comprehension of the software, and \textbf{4)} they unnecessarily increase the software size, thus increasing the memory footprint of the software and forcing more frequent hardware upgrades. Therefore, source code tools that identify exact and similar code fragments have become a common practice in software development for similar code identification \cite{roy2018benchmarks,rattan2013software}. However, they are constrained by the size of the code base to be searched, their ability to find duplicate code with minor modifications, and their ability to present detection results in useful and usable manner to development and quality teams.

Typically, source code similarity detection tools work by scanning the source code to identify pieces of code that are similar by using different string matching algorithms. 
Although reasonably fast for small data, they are quiet inaccurate and slow for large data.
Additionally, the code fragments can be textually different but share similarity at semantic or structural level due to modifications made to the code such as variable renaming, statements insertions, deletions, and replacements. 
Furthermore, precise (i.e., exhaustive) search of code fragments is often infeasible, therefore, the tools tends to over-approximate, so not to miss any possible duplicates.  
Consequently, this leads to generating a considerable number of false positives that needs to be manually inspected and verified \cite{tiarks2009assessment,juergens2010achieving}. Therefore, there is a need for source code search and classification techniques that can handle large source code repositories efficiently with reasonable accuracy to be useful for  development and quality teams.

In this paper, we present our approach for source code similarity detection in context of the SAP ABAP programming language. The presented approach is designed to be robust and scalable for large code base. We present our initial experimental results and experience in implementing a prototype of the approach.
The rest of the paper is organized as follow. Section 2 provides brief background including general terms, definitions and some related work. The proposed approach is introduced in Section 3. In Section 4 we discuss experiment settings and show some results. Lastly, we conclude the paper and present  future work in Section 5. 

\begin{figure}[!t]
\centering
\includegraphics[width=.4750\textwidth,height=.235\textheight]{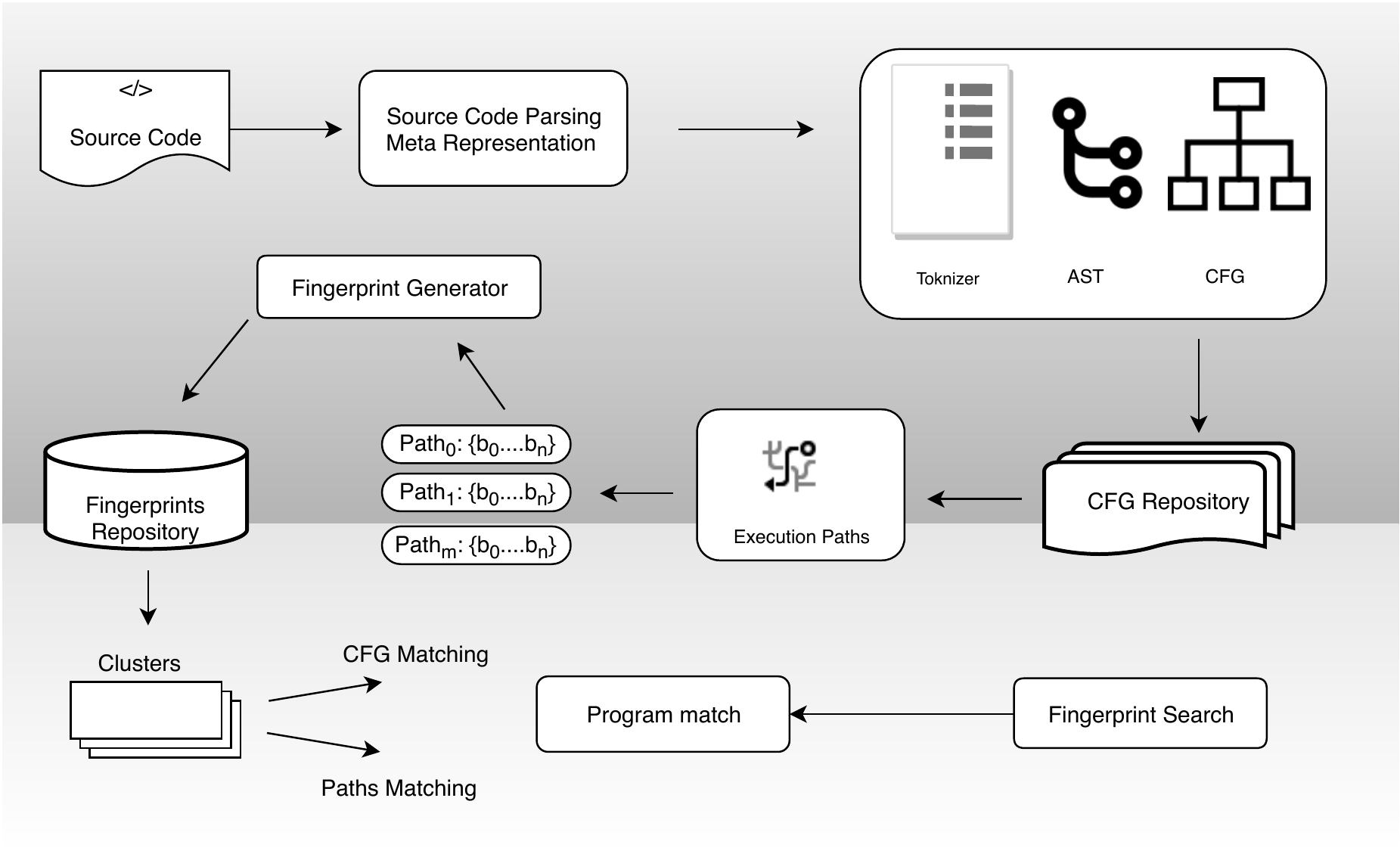}
\caption{Similarity Detection Overview: High level steps of the proposed approach.} \label{fig1}
\end{figure}

\section{Background and Related Work}

Clones can be broadly categorized into four types based on the nature of their similarity  \cite{roy2009comparison,rattan2013software,svajlenko2014evaluating,bellon2007comparison}. \textbf{Exact Clones (Type-1)}: are clone pairs that are identical to each other with no modification to the source code. \textbf{Renamed clones (Type-2)}: are clone pairs that are only different in literals and variable types. \textbf{Restructured or gapped clones (Type-3)}: are renamed clone pairs with some structural modifications such as additions, deletions, and rearrangement of statements. \textbf{Semantic clones (Type-4)}: are clone pairs that have different syntax but perform the same functionality (i.e., semantically equivalent). These typically are most challenging to find and identify, yet, they are the more relevant in the context of ERP systems \cite{thomsen2012clone,guo2008detecting}.

Several approaches have been proposed in the literature to identify similar source code ranging from textual to semantic similarity identification. Generally, they're classified based on the source representations they work with. 
In \textbf{Text based detection}, the raw source code, with minimal transformation, is used to perform a pairwise comparison to identify similar source code \cite{ducasse1999language}. 
\textbf{Token based detection} on the other hand, extracts a sequence of tokens using compiler-style source code transformation \cite{baker1995finding}. The sequence is then used to match tokens and identify duplicates in the repository and the corresponding original code is returned as clones. 
In \textbf{Tree based detection}, the code is transformed to Abstract Syntax Trees (ASTs) that are then used in tree sub matching algorithms to identify similar sub trees \cite{baxter1998clone}. Similarly, clone detection is expressed as graph matching problem for Program Dependence Graphs (PDGs) in \cite{krinke2001identifying}. % in graph matching algorithm to find similar code. 
\textbf{Metrics based detection} extracts a number of metrics from the source code fragments and then compare metrics rather than code or trees to identify similar code \cite{mayrand1996experiment}.

Generally, similar code identification techniques work at varying level of granularity. \textbf{Fine-grained} detection leverages tokens, statements and lines as the basis for detection and comparison \cite{kamiya2002ccfinder}. On the other hand, \textbf{coarse-grained} detection uses functions, methods, classes, or program files as the basic units of detection \cite{roy2009comparison}. Naturally, the finer the granularity of the tool is, the longer time it takes to find clone candidates. Equally, the larger the granularity of the tool is, the faster time it takes for detection, albeit with fewer detected clones \cite{sheneamer2015code}. Detection tools have therefore to make design trade-offs between accuracy and performance on an almost constant basis based on the code base being examined.

Another challenge in finding duplicate code is the performance of querying and retrieving possible matches from a large code base. Fingerprinting and hashing have been used to improve the search efficiency \cite{hummel2010index}. Hashing maps variable size source code to a fixed size fingerprint that can later be used to query and search for clones in linear time \cite{toomey2012ctcompare}.  However, a simple match doesn't work well for inexact matches. Others \cite{thomsen2012clone,uddin2011effectiveness} use hashing techniques to group similar source code fragments together, thus enhancing the accuracy and performance of clone detection techniques. However, this is less effective in detecting Type 4 clones as hashing and fingerprints are based on the source code and not its semantic. Machine learning approaches have been proposed \cite{white2016deep} to link lexical level features with syntactic level features using semantic encoding techniques \cite{peng2018bayesian} to improve Type 4 clone detection. However, in order  for them to be effective, human experts need  to analyze source code repositories to define features that are most relevant for clone detection. 

%However,  even  the enhanced  approach  with  expert-defined  features  [32]  cannot detect  vulnerabilities  that  are not caused  by  code  clones. However, this requires learning from large source code repositories.
%syntac machine learning techniques are used to improve Type 4 clone detection by

One way to capture the program semantic is the code Control Flow Graphs (CFGs) . CFGs are one of the intermediate code representations that describes in graph notation, all paths that might be followed  through a piece of code during its execution \cite{grove2001framework}. In CFGs, vertices represent basic blocks and edges (i.e., arcs) represent execution flow. 
Since CFGs capture syntactic and semantic features of the  code,  they are better at resisting changes in the code that manipulate source code in very minor ways, while not affecting the functionality of the program. For this reason, control flow graphs have been used in static analysis \cite{mikhailov2016control}, fuzzing and test coverage tools \cite{sparks2007automated}, execution profiling \cite{ball1996efficient,wu1994static}, binary code analysis \cite{lim2014comparing}, malware analysis \cite{bruschi2006detecting}, and anomaly analysis \cite{nandi2016anomaly}.
%\textit{ To the best of our knowledge this is the first work presenting a similarity detection for source code based on CFG hashing. }

In this paper, we argue that clone detection is characterized by more than just text patterns in the source code because, there is semantic features as well as syntactic features that must be considered for effective similarity detection. Since CFGs deliver both syntactic and semantic information of the code, we argue that the Control Flow Graphs (CFG) representation provides a sensible choice for source code similarity detection for the following reasons. First, CFG blocks boundaries represent an intrinsic granularity level that is neither too fine nor too coarse for clone identification. Second, CFG provide a reasonable balance between syntax and semantic representation of the code, thus, considering clones with more than text pattern similarities.

We, therefore, present a  clone detection technique for improving the precision of detecting similar code \textit{clones} with reduced time and memory complexities associated with large code base. The key idea is to avoid cost associated with graph and tree pattern matching by utilizing a staged identification approach. Specifically, we employ normalization and abstraction to standardize the source code. Then we derive CFG and enumerate a list of possible execution paths. Finally, context sensitive hashing is used to efficiently fingerprint the code for approximate similarity search and matching. Similarities between fingerprints can be done in linear time to identify a subset of the code base that is close to each other. Then a specific matching is performed on the subset that meets a certain similarity threshold. 
The presented approach addresses practical scalability complexities and give the flexibility to incrementally store and re-use the processed data to enable efficient search and matching for source code. 
The approach scalability is achieved by leveraging a similarity preserving hashing technique on a CFG execution path-level granularity to compactly represent fingerprints and reduces the number of comparisons. Additionally, our semantic-aware CFG based abstraction technique renders our approach resilient to minor modifications not affecting the semantics of the program

\section{Proposed Approach}

In this section, we describe the proposed three-step (staged) clone detection for source code as shown in Figure~\ref{fig1}. The first step involves extracting source code meta-data up to the CFG for the code being evaluated. The second step involves processing the CFG to generate the program fingerprints. The third step involves identifying the most related code groups in the repository followed by similarity check among the group members.

\subsection{Normalization and Abstraction}

\begin{table}[!t]
\caption{Normalization and Abstraction Rules.}\label{tab1}
\begin{center}
\begin{tabular}{ l l }
\hline
\textit{Pattern} & \makecell{\textit{Proposed Transformation}} \\
\hline
\textbf{Local Variables} &	 \makecell{ \textit{L-Variables}}\\
\textbf{Global Variables} &	\makecell{\textit{G-Variables}} \\
\textbf{loops: For, Do, while} &	 \makecell{\textit{ Iterate <start>} \\\textit{ <condition> <+/->}} \\
\textbf{Conditions: If, else, case} &  \makecell{{\itshape Selection <condition>}}\\
\hline
\end{tabular}
\end{center}
\end{table}

Code normalization is the process of transforming a piece of code to remove all the irrelevant parts of the code for the comparison. Applying normalization and abstraction to the code increases the clone variations that can be detected. This includes removing comments, white spaces, empty lines which don't affect the program behavior. Literal values, identifier names are fixed with specific tokens. Abstraction structures such as \textit{Loop}s, \textit{If}s, and \textit{Case} statements are also normalized to increase resilience against syntactic variations. Specifically, the lexer (i.e., tokinizer) is used to break the stream of code to tokens. The parser then generates the Abstract Syntax Tree (AST) from the tokens using the context provided by the language grammar. The code can then be normalized according to predefined rules such as the example ones provided in Table~\ref{tab1}.

One should note that the purpose of the normalization and abstraction step is to make the matching more resistant to semantically irrelevant variation. Which information to include or exclude is dependent on the specific language and on which kind of clones should be found. However, excessive normalization can introduce ambiguities that decreases the accuracy of the match. Therefore, it is important to carefully consider the level of normalization based on the envisioned use cases.

\subsection{Control Flow Graphs (CFGs)}

CFGs describe the order in which code statements are executed as well as conditions that need to be met for a specific path of execution to be taken \cite{allen1970control}. They capture the structure of a program by a directional graph (i.e., digraph) in which nodes (i.e., vertexes) define the program basic blocks and edges define the possible control transfers between these blocks. Specifically, a basic block is a continuous sequence of statements that executes in the same order as they appear in the block without control changes  (i.e., branches and jumps). Directed edges on the other hand,  represent jumps and branches between CFG basic blocks. 
The CFG construction is carried out based on the abstract syntax tree (AST) representation to which control flow information are introduced \cite{krinke2001identifying,mayrand1996experiment}.

\subsubsection{Preliminaries} We define a program $\varpi$ with $K$ statements $S=(s_0\dots s_k\dots s_K)$ as a set of basic blocks $B=(b_0\dots b_n\dots b_N)$.
The \textbf{basic-block} $b_i=(s_i \dots s_j)$ is sequence of statements without branch or control statements.
The \textbf{CFG} is a directed graph $CFG=(B,E)$, where $B$ specifies the basic blocks (i.e., nodes). $E \subseteq B \times B$ is the set of directed edges (i.e., control transfers), where $e_{i,j}=(b_i,b_j) \in E$ iff there is a control transfer (i.e., branch or a jump) from $b_i$ to $b_j$. Given $b_0, b_N \in B$, a \textbf{path} from $b_0$ to $b_N$ is a sequence of blocks (i.e., graph nodes) $p_r=(b_0\dots b_m \dots b_N) \in B$, such that there is an edge $(b_m,b_{m+1}) \in E$. The set of all possible execution paths in a CFG can then be expressed as $\mathcal{P(CFG)}=(p_0\dots p_r\dots p_R)$.

\subsubsection{Control Flow Extraction}
\label{CFE}
The steps to extract the CFG can be explained with the help of the pseudo code show in algorithm \ref{alg1_cfg}.
CFG extraction starts with the normalized source code obtained from the corresponding AST. A syntax tree represents the design with a tree structure by abstracting the details concerning the syntax of the language. This is then used to get the program normalized statements. These statements are then analyzed to identify labels to the code blocks called leaders. We identify labels for the code blocks (i.e., nodes) as follows: 1) the first statement of a program is a leader, 2) the targets of control statements (i.e., loops and conditions) are leaders, and 3) the statements immediately following control statements are leaders. The sequence of statements between these leaders constitutes the basic blocks $b_i \in B$. An edge $e_{i,j} \in E$ describes the transfer of control between two blocks of code $b_i$ and $b_j$. 
The CFG is constructed by adding edges between basic blocks, where execution control-flow exist.

\begin{figure}[!t]
\centering
\includegraphics[width=.480\textwidth,height=.25\textheight]{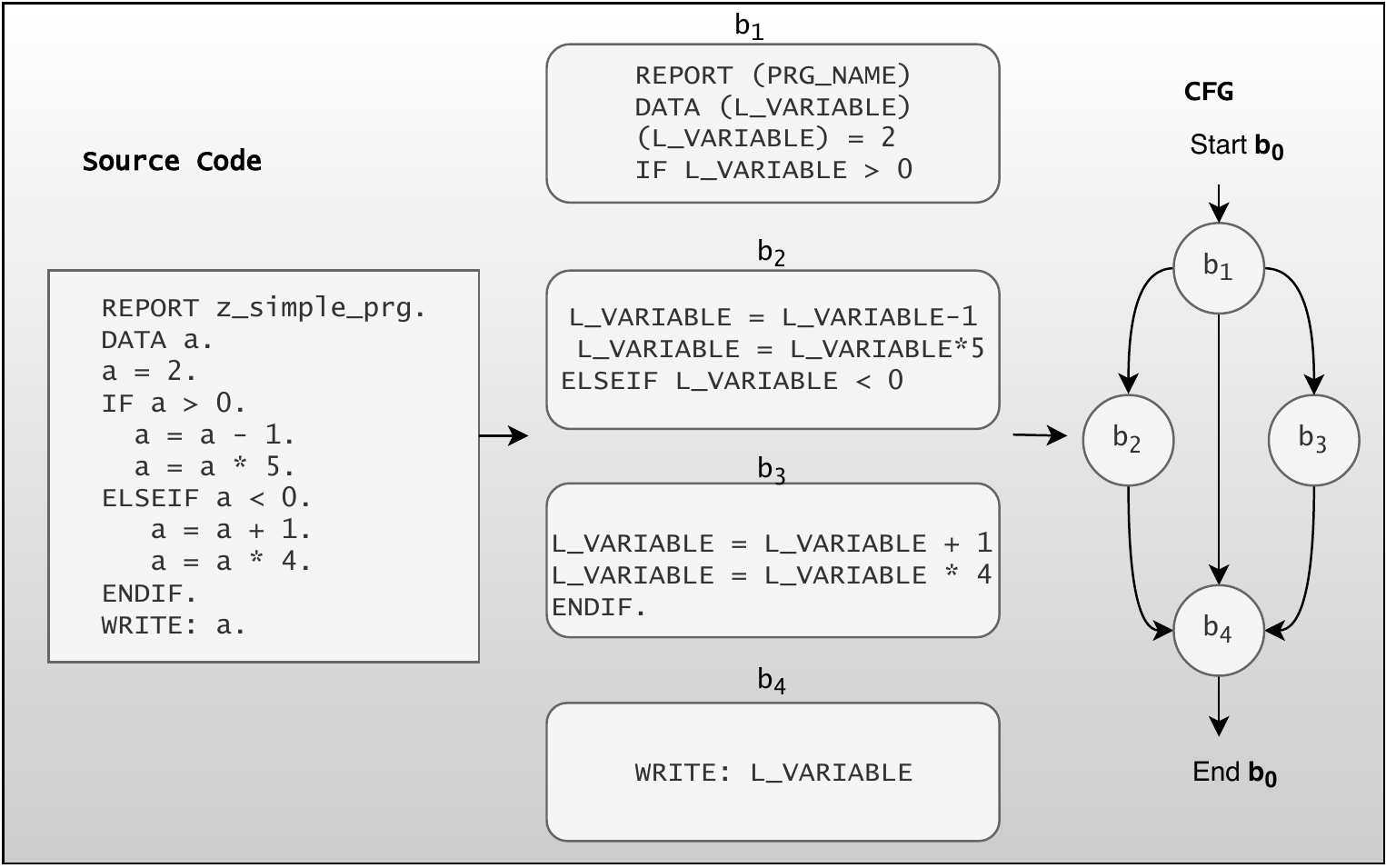}
\caption{Control Flow Extraction Example.} \label{fig2_stc}
\end{figure}

Figure~\ref{fig2_stc} illustrate how a CFG is extracted from a piece of example source code. It should be noted that the CFG extraction here is a simplified approach and doesn't consider recursive calls, function calls, or try catch statements at this stage. If more precise CFGs are deemed necessary one can use more advanced techniques, however, at the expense of the CFGs extraction time.

\begin{algorithm}[!t]
%\SetAlFnt{\tiny}
\KwIn{Normalized Source Code }
\KwData{Statements $S=(s_1,...,s_k,...,s_K)$ }
\KwResult{ $CFG=<B,E>$ }

\SetKwBlock{GetLs}{Get Leaders:}{}
\SetKwBlock{GetBlks}{Get Blocks:}{}
\SetKwBlock{GetCfg}{Build CFG:}{}

\nl \GetLs
{%getleaders
Leaders = $ \varnothing$ \;
Leaders $\gets s_0$ \;
\ForEach{$s_{k; k=(1,..,K)} \in S$}
{
\If{$s_k  \in  (isControl  \land  \neg isLast) $}
{ 
leaders= leaders.Add($s_{k+1}$) \; 
leaders = leaders.Add(target($s_k$))\;}
leaders= leaders.Add($s_k$) \; 
}%for
}%leaders

\nl \GetBlks
{
BasicBlocks \;
\ForEach{ $l$ in leaders}
{ $i=l$\;

 \While{($i < K)  \land \neg isLeader$ }
 {  
 $i=i+1$\; 
 }
 $i=i-1$\;
$b_k$=BasicBlocks.create($leader, i$)\;

}
}
%\Return $B$ \;
\nl \GetCfg
{
\ForEach{$b_n \in B$} 
{
$leaders \gets leaders - b_b$ \;
\ForEach{$s_k \in b_n$}
{
\If{$s_k \in isControl$}
{
$b_i=statementTarget(s_k)$\;
edges.Add$(b_n,b_i)$\;
}
}

}%forEach
\Return $CFG(p)$ \;

}

\caption{CFG Extraction Algorithm}
\label{alg1_cfg}
\end{algorithm}

\subsection{Execution Paths}
\label{EFE}

\begin{algorithm}[b]
\SetAlgoLined
\KwIn{$CFG=<B,E>$ , $b_0$, $b_N$ }
\KwResult{$\mathcal{P(CFG)}=(p_0,\dots,p_L)$ }
\SetKwFunction{getPaths}{\textit{\textbf{Paths}}}
\getPaths{$b_{source}, b_{destination}, path, Visited$}\;
\Indp

Visited($b_{source}$)=true\;
%$Visited \cup \{b_{source}\}  $\;

\If{$b_{source} = b_{destination}$}
 {
 %PathsList.add(path)\;
 $\mathcal{P(CFG)} \gets \mathcal{P(CFG)}  \cup \{ path \} $\;
 Visited($b_{source}$)=false\;
 \Return
 }
\ForEach{successor $\in b_{source}$.successors() }
 {
 \If{$ \neg $ Visited($b_{successor}$)}
 { %$p_l = p_l \cup \{b_{successor}\}$\;
 path.Add(successor)\;
 \getPaths{$b_{successor}, b_{destination}, path, Visited$}\;
 path.Remove(successor)\;
 }
  }
 Visited($b_{source}$)=false\;
 \Return %$\mathcal{P(CFG)}$
 \caption{Path Enumeration}
 \label{alg3_paths}
\end{algorithm}

Execution paths model different possible execution order of program statements, which contain many loops, exceptions, and calls; that is to say, features that reflect the semantics of the program. More precisely, the execution paths consider the nodes and the inter-dependencies (i.e., edges) between different nodes in the CFG of a program. Therefore, they can reproduce some of the semantic effects of a particular CFG.

Given the CFG obtained in section \ref{CFE} one can define all possible execution paths of a program such that every two adjacent nodes in a path are connected by an edge in $E$, where $b_0$ is the start block and the end node is $b_N$. For any given program run only one path, among all the possible paths, can be followed. Therefore, specific program behavior can be viewed as collection of these paths. Pseudo code in algorithm~\ref{alg3_paths} enumerate all potential paths using a depth-first traversal of the CFG starting at $b_0$. Each path is a stack of basic blocks and each block $b_n$ is a sequence of statements $s_k$. A program CFG can then be expressed as the set of possible execution paths $\mathcal{P(CFG)}$ for the program CFG. 

One should note that the paths enumeration technique presented here is not precise. However, we argue that exact paths are not important in themselves. Rather, what is important is that these paths are used to provide additional execution context to the CFGs. We can therefore, use paths with lower accuracy, yet still acceptable for the purposes of our approach.

Once the all potential paths are enumerated, the fingerprints can be created for each path as described in the next section.

\subsection{Fingerprint Generation}

CFGs fingerprinting involves extracting various features from the CFGs. The features are used to generate a unique and compact representation of the CFGs in such a way that similar paths are assigned similar fingerprints. Typically, the extracted features can be either semantic or syntactic features, which are then hashed to produce the compact representation of the CFGs. However, on the one hand, using syntactic only features will result in a fingerprint that is sensitive to minor code modifications. On the other hand, using semantic only features will result in a fingerprint that is sensitive to CFG structural information, ignoring block statements syntactic features. Therefore, our objective is to generate a fingerprint that capture not only syntactic information of a CFG, but also semantic features of a CFG as well.

We make use of the extracted execution paths in section \ref{EFE} to represent the CFGs in a light-weight fingerprint. The fingerprint, in addition to compactly capturing textual and structural information of the programs, should represent each path in the program, such that similar paths have a higher probability of collisions or will only differ slightly in their digest. Specifically, very similar paths should map to very similar, or even the same, digest, and difference between digests should be some measure of the difference between paths.

To this end, we employ a similarity preserving hashing techniques \cite{wang2014hashing, wang2018survey, datar2004locality} to fingerprint the CFGs. Our choice is motivated by the following reasons: \textbf{1)} fingerprints shorter size lend themselves well to efficient search and clustering algorithms, thus speeding search time; and \textbf{2)} similarity preserving fingerprints incorporates approximation, thus capturing more clones than a strict text or graph isomorphism-based approaches.

With these observations in mind, we extend the SimHash technique presented in \cite{charikar2002similarity} to generate a  fingerprint that is both efficient and also considers the program execution semantics as captured by the control flow graph. 
Concretely, all sequence of blocks along a certain execution path are stacked as one unit. The sequence of statements in these ordered blocks are then hashed as shown in in algorithm \ref{alg2_fp}. Therefore, an execution path will have a specific hash value and a program CFG will have multiple hash values.

\begin{figure}[!t]
\centering
\includegraphics[width=.25\textwidth,height=.135\textheight]{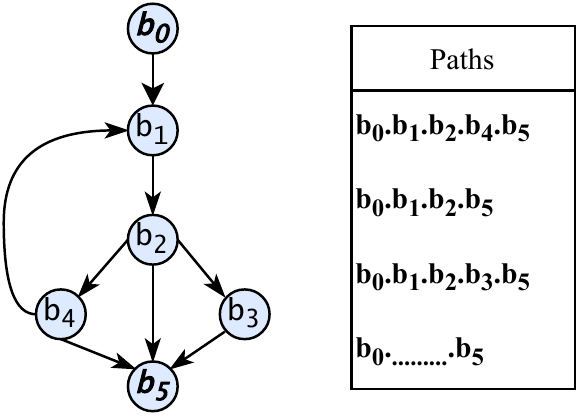}
\caption{An Example Control Flow.} \label{fig2_hash}
\end{figure}

Consider the CFG example shown in Figure~\ref{fig2_hash}. The possible paths for the program $\varpi$ are the set of paths in $\mathcal{P(CFG)}=( p_0=<b_0,b_1,b_2,b_4,b_5>, p_1=<b_0,b_1,b_2,b_5>, p_2=<b_0,b_1,b_2,b_3,b_5>)$. Accordingly, program $\varpi$ fingerprint \[\mathcal{H(CFG)}=(h(p_1),..,h(p_m)..,h(p_M)),\] is a set of CFG paths hash values, where $h(p_m)$ is computed as shown in algorithm \ref{alg2_fp}. Specifically, basic blocks statements are hashed by a traditional hash function. We use \textit{Murmur64} hash in our implementation, however, any other hash can be used. A vector $\vec{v}=(v_1, \dots ,v_R)$ is initialized to zeros. $R$ is selected based on the desired hash size in bits. The algorithm scans all statements in the block, $v(r)$ is incremented by a weight $w=1$ if the $r^{th}$ bit of the statement is one, otherwise, $v(r)$ is then decremented by one. Then,  $h(p_m)_r$ bits are set to one if $v(r)$ is not zero. Finally, $h(p_m)_r$ is added to the program fingerprint $\mathcal{H(CFG)}$.

Observant reader may argue that it would be redundant and space consuming to consider information captured from repeated blocks that are common in the multiple paths of the same CFG. However, our fingerprint is compact and computed efficiently thus making repetition impact negligible. Moreover, further refinement and fine-tuning are also possible to exclude irrelevant paths and blocks from the fingerprint.   

\begin{algorithm}[!t]
\SetAlgoLined
\KwIn{$\mathcal{P(CFG)}$, $\vec{v}=(v_1,..,v_r)$}
\KwResult{$\mathcal{H}(CFG)$ }
initialization\;
\ForEach{ $p_m \in \mathcal{P(CFG)} $ }
{

$\vec{v} \gets 0$\;

\ForEach{$b_n \in B$} 
{
\ForEach{$s_k \in b_n$}
{
$h^s(s_k)=$ MurmurHash64($s_k)$\;
\For{ $i=0  \rightarrow R$}
{
\uIf{$i^{th}$ bit of $h^s(s_k) == 1$ }
{$v[i]=v[i]+w$\;}
\Else{ $v[i]=v[i]-w$\; }
}%for i
}%for S 
}%for B
$h(cfg) \gets 0$\;
\For{ $i=0  \rightarrow R$}
{
\If{$v[i]>1$}{$h(p_r)_i=1$\;}
}%for R
$\mathcal{H}(CFG)= \mathcal{H}(CFG) \cup \{h(p_r)\}$\;
} 
 \caption{Control Flow Graph Fingerprinting}
 \label{alg2_fp}
\end{algorithm}

\subsection{CFGs Matching}

Once the fingerprint for the CFGs is generated, the actual matching is performed. For CFGs to be considered similar they not only have to be isomorphic, but also the basic blocks of CFGs has to match. Specifically, graph matching is performed by checking node similarity, edge similarity, and the relationships between them. One can make use of graph matching techniques such as bipartite matching and maximum common subgraph isomorphism (MCS) \cite{raymond2002maximum}. However, most of graph matching problems are NP-complete and suffers from a high computational complexity. 

One can also use the CFG fingerprints to compare pairs of CFGs. However, pairwise exact matching is neither efficient for large repositories, nor robust enough to match programs with small variations. Therefore, we use inexact matching of the fingerprints. Specifically, in our case, once we get all the fingerprints, the problem of detecting code clones becomes essentially a fingerprints categorization and clustering problem, such that, in a cluster, the pair-wise a similarity metric remains below a pre-defined threshold value, $\alpha$ , while restricting the cluster size to be no less than another pre-defined value.
This is accomplished with the help of a similarity function which is described next. 
 
\subsubsection{Similarity Function}

The idea of our similarity function is to have a measure of structural similarity of two CFGs which not only looks at the CFG structure, but also on the meta-data of the these nodes. For example, a pair of CFGs may be isomorphic (i.e., identical), however, the nodes (i.e., blocks syntax) are different. Another pair of CFGs may not be isomorphic but have similar paths. Therefore, CFG comparison should consider the path basic blocks similarities as well as the actual paths common between two CFGs.  

We estimate the similarity between a pair of paths $(p_i,p_j)$ as the number of bits which differ between the two fingerprints. More formally, given two fingerprints $h(p_i)$ and $h(p_j)$, expressed as a binary vector of length $R$, we define the distance $p_i$ and $p_j$, $D(p_i,p_j)$, to be the number of bits where $h(p_i)$ and $h(p_j)$ differ. The lower the value of the distance, the more similar the paths are. For example, a distance value of 0 means that the paths are identical, while a distance value of $R$ means that the two paths are dissimilar.   
The $\mathcal{S}_{paths}=(D_{path}(p_i,p_j) :  \forall (p_i,p_j) \in CFG \times \overline{CFG})$ denotes the set of pairwise comparison values (i.e., hamming distance) between paths in two distinct CFGs.

Earlier work \cite{manku2007detecting} shows that one can efficiently identify whether fingerprint pairs differ in at most $\alpha$ bits. This value can be seen as a threshold for similarity between two fingerprints. Specifically, the lower the value of $\alpha$ the higher the similarity between the path blocks. Furthermore, different values of $\alpha$ represent different degrees of similarity. For example, $0< \alpha <4$ represent identical or near identical clones, while $4< \alpha <8$ represent similar but not near identical clones. In our approach we use $\alpha<8$ empirically based on our experimental findings, however, other values can be used for different environment settings.           

The term $\mathcal{S}_{CFG} \in [0,1]$ computes the pairwise similarity between two CFG fingerprints. A similarity value of 1 means that the programs are similar, while a value of 0 means that the two programs CFG share no similar paths in common. Specifically, we use a variant of the Jacard index to estimate the overall similarity between the pair $(CFG, \overline{CFG})$ is then defined as: 
\begin{equation}
\mathcal{S}_{CFG}(CFG, \overline{CFG}) = \dfrac{|\{s \in \mathcal{S}_{paths} : s \leq \alpha \}|}{\min (|P(CFG)|,|P(\overline{CFG})|)}.
\end{equation}
The numerator is the number of common paths with at most $\alpha$ different bits and the denominator is the total number of paths in the smaller program CFG. Consequently, the overall similarity between the two programs can be seen as the degree of overall similarity between similar paths between two the programs.
  
%Practically, one can use the fingerprints computed for each program CFG to classify the source code repository into several groups according to the similarity of the fingerprints. To find an initial match, we identify categories that are close to the fingerprint. Categories that are the farthest can then be excluded to reduce the the number of comparisons required to find an initial match. Moreover, the distance between two fingerprint can be further improved by using bit parallelism and bit-counting optimization.

One should note that the term $\mathcal{S}_{CFG}(CFG, \overline{CFG})$ computes when one CFG is contained inside of another. In other words, it considers the case where one program consists of repeated copies of another smaller program. If we want to measure the total amount of resemblance, that is proportional similarity, between two programs CFGs one can change the numerator to be the total number of common paths. However, with either measures one can use $\alpha$ to only consider candidates where the similarity or containment score meets a pre-determined threshold.

\section{Evaluation And Discussion}
In this section we present the initial results of our proof of concept along with some insights gained during our evaluation. First, we present the details of the experimental setup and the data selected for the evaluation. Then, we present the results and discuss different observations made during the experiment. 

\subsection{Experiment}
To evaluate the effectiveness of our approach in detecting similarities in source code, we ran the evaluation on a synthetic data set. We assessed the performance of the proposed approach against two criteria. First, execution time efficiency, and second, detection precision. 
The detector is implemented using Java. Specifically, the detector reads the source code files and performs normalization and generates the control flow graphs. The control flow graphs are represented using the DOT format where nodes contain the normalized source code of a block and edges connecting the blocks. All the experiments were conducted on a Windows 10 computer  with  an  Intel  Core  i7  (4 cores at 3.60GHz) processor, 16GB of RAM and 1TB of HDD storage capacity. 

\subsubsection{Experimental Data}
Due to the lack of a standard benchmarks for ABAP clone detection, 664 ABAP programs were collected from online repositories\footnote[1]{https://github.com/trending/abap} to serve as our code base. Typically, Verifying clones is a subjective decision that depends on the analyst experience and the context of the code. Furthermore, manual verification of all code clones candidates is impractical. Therefore, We selected a 50 random programs for manual examination to evaluate the existence of false positives.

\subsection{Results}

\begin{figure}[!t]
\centering
\includegraphics[width=.47750\textwidth,height=.275\textheight]{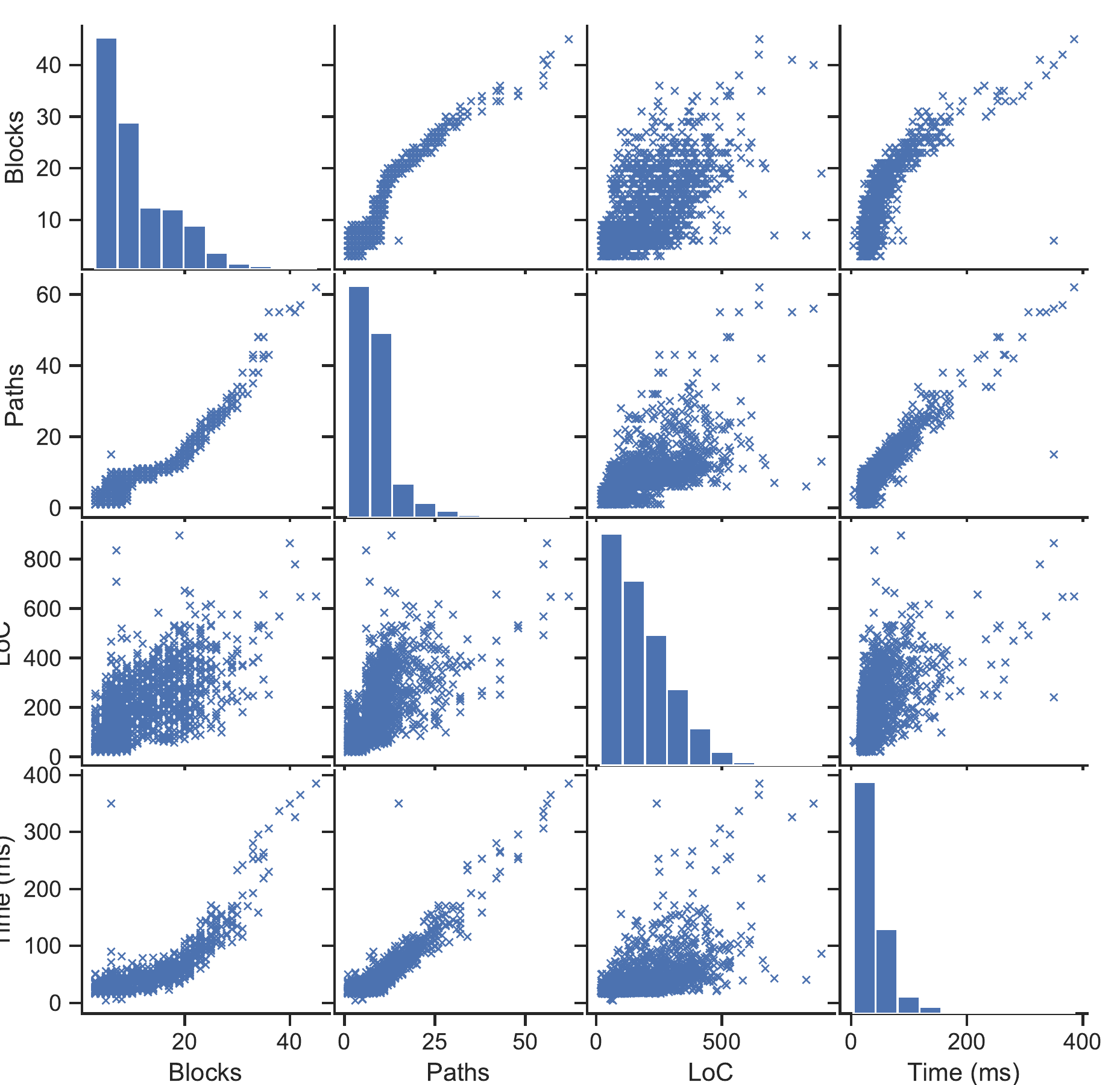}
\caption{Pairwise plot for the experiment data set} \label{fig_pwp}
\end{figure}

\begin{figure*}[!ht]
     \centering
          \begin{subfigure}[b]{0.245\textwidth}
         \centering
         \includegraphics[width=\textwidth, height=3.5cm]{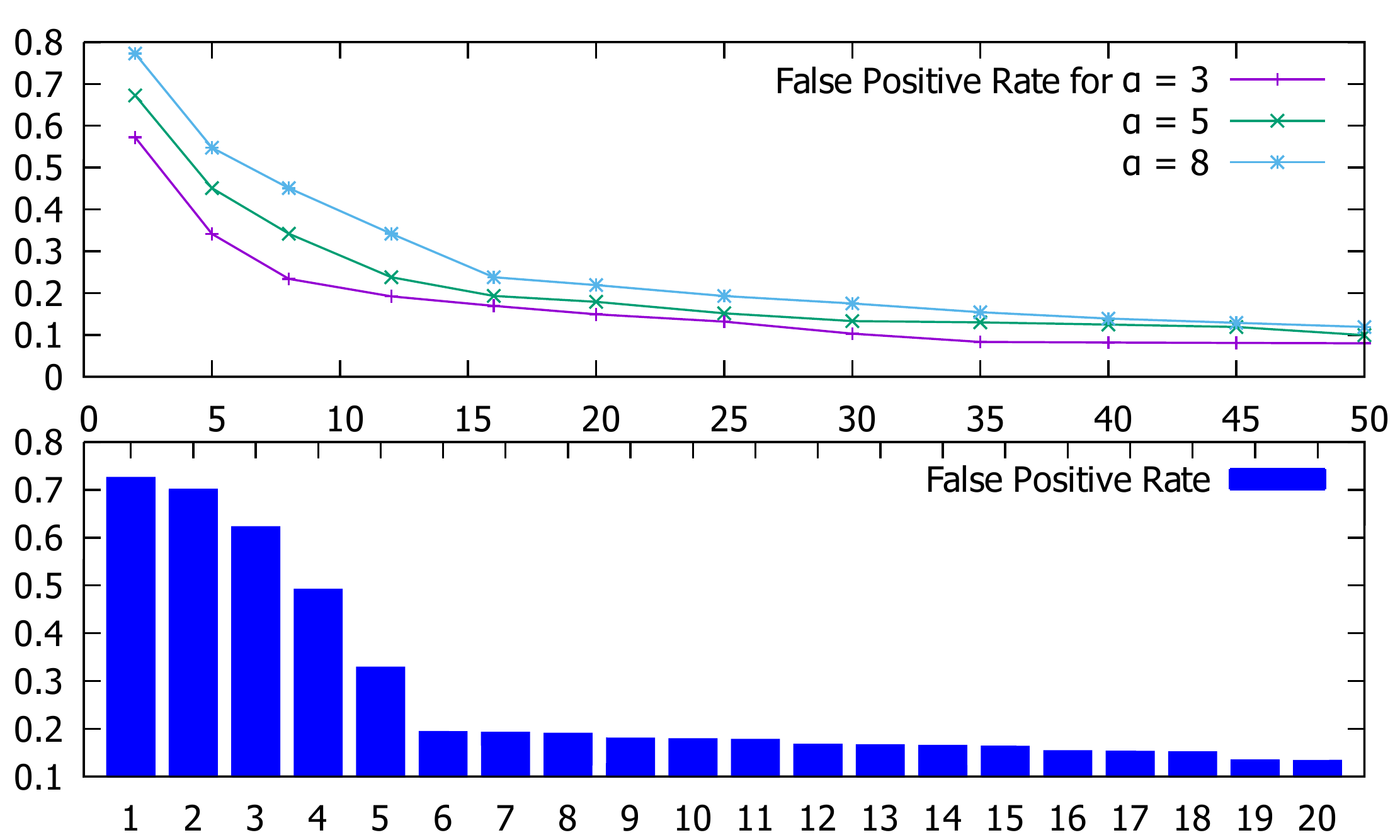} 
         \caption{}
         \label{fig:FPR}
     \end{subfigure}
     \hfill
     \begin{subfigure}[b]{0.245\textwidth}
         \centering
         \includegraphics[width=\textwidth, height=3.5cm]{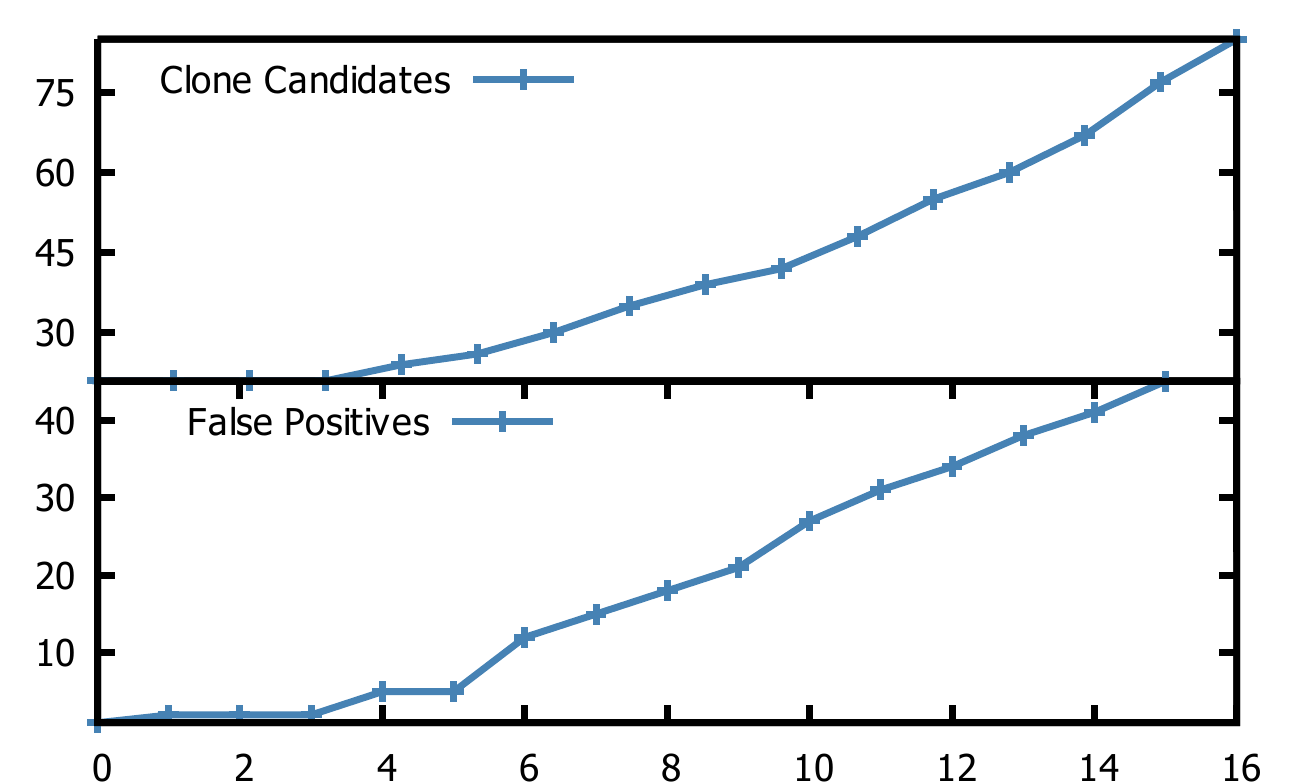}
         \caption{ }
         \label{fig:Accuracy}
     \end{subfigure}
     \hfill
     \begin{subfigure}[b]{0.245\textwidth}
         \centering
         \includegraphics[width=\textwidth, height=3.5cm]{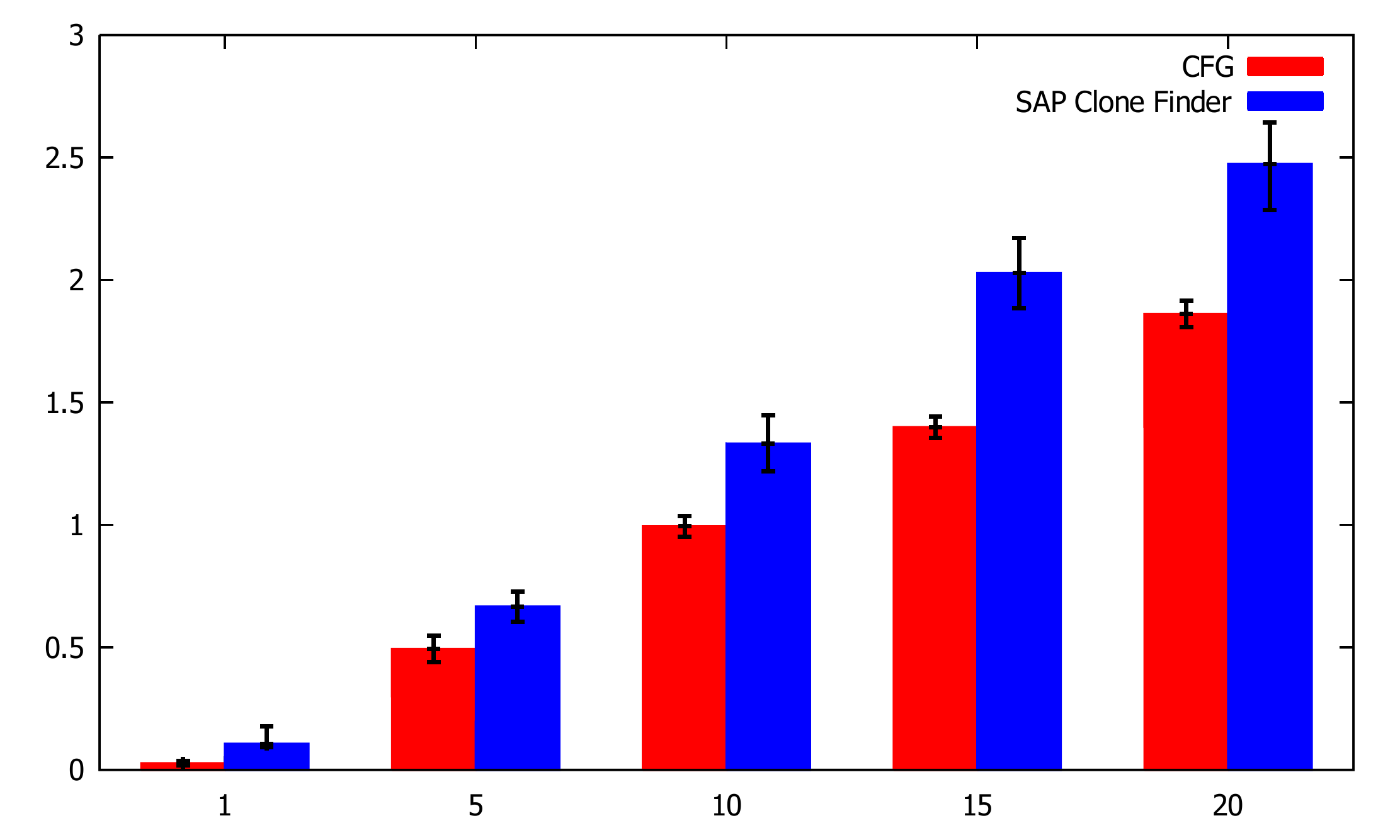}
         \caption{}
         \label{fig:RT}
     \end{subfigure}
     \hfill
     \begin{subfigure}[b]{0.245\textwidth}
         \centering
         \includegraphics[width=\textwidth, height=3.5cm]{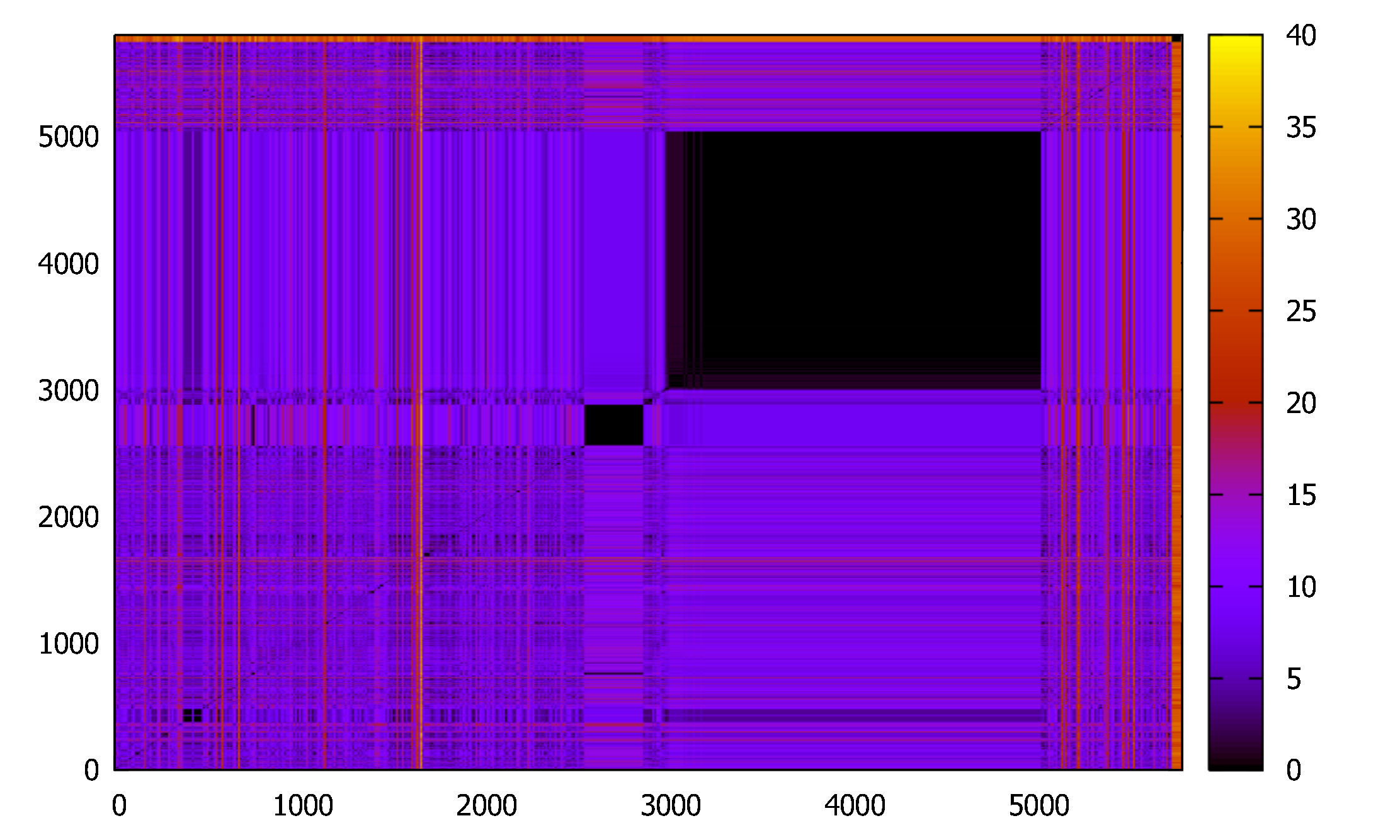} 
         \caption{}
         \label{fig:Sim}
     \end{subfigure}
     
        \caption{Detection Results: (a) False positives as a function of the number of LoC (Top), False positives as a function of the number of blocks (Bottom); (b) Clone candidates identified for $\alpha$ (Top), False positives for $\alpha$ (Bottom); (c) Time it takes to query for clones in our approach; and (d) Similarity Matrix Heat Map. }
        \label{fig:three graphs}
\end{figure*}

Figure~\ref{fig_pwp} shows the scatter plot of the source code set used in our experiments. The pairwise plots show the relationships between the number of blocks, paths, line of code, and the CFG extraction time for the programs in the data set. One can see that most of the programs in the data set contains between 5-20 blocks, that results in less than 25 paths for most of the programs in the data set. The number of blocks and paths are closely correlated, however, the number of lines in the program is less correlated to the number of path. Furthermore, the execution time grows as a function of the paths, blocks and to a lesser extent line of codes. On average the execution time was 0.17 seconds and  all cases the execution time didn't exceed 0.45 seconds. 

The distribution of the blocks count in programs show that a considerable number of programs with 10 or less blocks and paths with less than 3 blocks. Intuitively, programs and path with lower number of blocks are less diverse and will generate a less precise fingerprint. This can be illustrated with the help of figure~\ref{fig:FPR}.  The figure shows that the precision increases as the number of basic blocks increases. Furthermore, examining paths that have only 1 block revealed that most of these blocks contain generic code related to language specific style requirements such as error handling, calls, and GUI related. This code is not necessarily useful for the purposes of clone detection. Therefore, we exclude all paths that contain less than 3 blocks from our fingerprints repository. 

Similarly, the precision increases as the number of lines of code in the block increases as shown figure~\ref{fig:FPR}. Additionally, smaller threshold $\alpha$ is slightly more accurate for smaller programs, however, this improvement becomes less pronounced as the block size increases. Another observation is the the value of $\alpha$ is sensitive to the size of the program. For example, larger program can tolerate higher values of $\alpha$ with a reasonable false positives, while smaller programs will provide a more accurate results with lower values of $\alpha$. 

The impact of the threshold $\alpha$ on the number of clones candidates identified and false positives is shown in Figure~\ref{fig:Accuracy}. The detector is able to identify clones with no false positives when $\alpha <= 4$. The false positives are still reasonable for  $\alpha <= 7$ , however, they become more pronounced at higher thresholds. This can be explained by the fact that at higher $\alpha$ means there are more bit differences between the signatures. We chose $\alpha=5$ as reasonable threshold for our experiment. The detector is able to identify most of the relevant clones with a reasonable false positives. larger $\alpha$ will identify more clones candidates, however, most of the newly identified one were false positive. Smaller $\alpha$ only detected near-identical clones and missed many programs.

Figure~\ref{fig:RT} illustrates the performance improvements compared to the standard clone detector in SAP systems. One can notice that our detector executes in linear time and grows with the number of programs under review. Nevertheless, our detector provides statistically significant performance improvements as shown in the figure. Moreover, SAP clone finder grows at faster rate with the number of programs being converted. The number of clones being detected is also significantly higher in our approach. This can be explained by the flexibility in our fingerprint representation and the value of $\alpha$.  

Figure~\ref{fig:Sim} shows the distance heat map (i.e., the values for the similarity matrix) for our code. The paths that are identical will have a lower distance, thus, darker colors. For example, There are quit few paths that are identical and therefore, shown in black. The structure shows that our method is able to discriminate between identical (or near identical) paths and similar paths. Furthermore, it validates our choice for $\alpha$ in the experiment. 

\subsection{Discussion}

While we have demonstrated the effectiveness of our approach for source code similarity search, there are still several practical issues one have to carefully consider for large scale adoption. Mainly, in performance and accuracy.

First, in very large code repository search and detection time may not be acceptable even with the improvements provided by our approach. However, one of the salient features of our CFGs fingerprint it's ability to represent group similarities. For example, one can use the fingerprints computed for each program CFG to classify the source code repository into several groups according to the similarity of the fingerprints. To find an initial match, we identify categories that are close to the fingerprint. Categories that are the farthest can then be excluded to reduce the the number of comparisons required to find an initial match. This can be done as an initial step to identify possible initial matches.

Another performance improvement is leveraging bit parallelism and bit vector arithmetic for the distance and similarity computations. For example, in \cite{jang2011bitshred} the Jaccard index is approximated using bit vector counting optimization. Same concept can be applied to our Similarity measure. Similarly, \cite{manku2007detecting} showed how one can find fingerprints with a certain threshold $\alpha$ (i.e., hamming distance) efficiently.

One should also note that the accuracy and precision of our approach relies on the CFGs extraction and path enumeration techniques. Naive techniques may result in more false positives. However, more robust techniques may be computationally expensive. Therefore, they need to be considered carefully. One can leverage different optimization techniques and use self tuning methods \cite{alomari2013self,alomari2012autonomic} to strike a balance between these conflicting requirements. Operationally, our tool may offer several static analysis techniques out of the box that may be selected and fine-tuned for the specific environment requirements.   

Another challenge is the fact that inserting jumps that are never taken in the CFGs distorts the CFGs by generating paths that will never be executed. While matching based on CFGs semantics is not possible in the general case, static profiling techniques such as the ones presented here \cite{ball1996efficient,wu1994static} can be used to check for branching and path probabilities. This can improve the semantic matching accuracy of our approach.

Furthermore, the size of the CFGs is a critical factor as shown in our experiment. We excluded the smaller CFGs blocks from our signatures. We believe this is a reasonable assumption since smaller code have significantly lower chances to be copied. However, the threshold should be investigated further. 

Using fine grained semantic at the block level to reveal more features is another area that is worth exploring. 
While some of this can be handled by fine-tuning the normalization and abstraction step in our approach, it might be helpful to explore some graph theoretic and machine learning ideas \cite{peng2018mining,peng2018block} to further improve accuracy. This however may increase performance requirements and should be considered as an additional refining step.

\section{Conclusion and Future Work}

This paper demonstrated the need to efficiently identify and measure source code syntax and semantic similarities in large code base of ERP systems and presented an approach to address this need. We designed a detection approach that searches for duplicate and near duplicate code in an efficient way. By leveraging similarity hashing to concisely represent the control flow graphs of the code, the fingerprints capture the programs intrinsic characteristics. The detector uses the control flow graphs to enumerates possible execution paths and fingerprint these paths efficiently. The experiment showed the viability of our approach and illustrated how the detector can achieve reasonable accuracies efficiently compared to current tools.

While the results of presented in the evaluation look promising, they present an initial results of our ongoing research in clone detection for large systems and there is considerable work to be done. For example, in addition to continuing our empirical validation for larger and more challenging code base, we plan to continue our work in several important directions. First, the CFG extraction can be enhanced to capture exception handling, and function calls in the representation. 
Since SAP systems are highly integrated systems, in  which data-centric programming is carried out in ABAP, one may also consider evaluating other intermediate representation such as call graphs, data flow graphs and system dependency graphs to consider the data-flow dependencies as well as control flow dependencies.

Secondly, our naive path enumeration can be improved to consider dominance relationship, loops and back edges to provide more precise paths for the CFG. It might also be worthwhile exploring the possibility of using path and execution static profiling techniques and path execution frequencies in the fingerprints to improve expressiveness of our representation.

The similarity hashing used can be also improved by exploring more intricate weights $w$ for the most relevant parts instead of the equal weights used in the current implementation. 
Other similarity hashing techniques can be explored and studied. Finally, we plan to apply our approach on industrial case studies to evaluate different practical considerations, which would provide more insight into scalability and usability questions for different situations.

\bibliographystyle{vancouver}

\begin{thebibliography}{10}

\bibitem{kremers2000enterprise}
Kremers M, Van~Dissel H.
\newblock Enterprise resource planning: ERP system migrations.
\newblock Communications of the ACM. 2000;43(4):53--56.

\bibitem{lee2003enterprise}
Lee J, Siau K, Hong S.
\newblock Enterprise Integration with ERP and EAI.
\newblock Comm of the ACM. 2003;46(2):54--60.

\bibitem{themistocleous2001erp}
Themistocleous M, Irani Z, O'Keefe RM, Paul R.
\newblock ERP problems and application integration issues: An empirical survey.
\newblock In: Proceedings of the 34th Annual Hawaii International Conference on
  System Sciences. IEEE; 2001. p. 10--pp.

\bibitem{brehm2001tailoring}
Brehm L, Heinzl A, Markus ML.
\newblock Tailoring ERP systems: a spectrum of choices and their implications.
\newblock In: Proceedings of the 34th annual Hawaii international conference on
  system sciences. IEEE; 2001. p. 9--pp.

\bibitem{keller2010official}
Keller H, Th{\"u}mmel WH.
\newblock Official ABAP Programming Guidelines.
\newblock Galileo Press; 2010.

\bibitem{keller2003abap}
Keller H, Kr{\"u}ger S.
\newblock ABAP objects.
\newblock Sap Press; 2003.

\bibitem{juergens2009code}
Juergens E, Deissenboeck F, Hummel B, Wagner S.
\newblock Do code clones matter?
\newblock In: Software Engineering, 2009. ICSE 2009. IEEE 31st International
  Conference on. IEEE; 2009. p. 485--495.

\bibitem{gupta2018survey}
Gupta A, Suri B.
\newblock A survey on code clone, its behavior and applications.
\newblock In: Networking Communication and Data Knowledge Engineering.
  Springer; 2018. p. 27--39.

\bibitem{roy2018benchmarks}
Roy CK, Cordy JR.
\newblock Benchmarks for software clone detection: A ten-year retrospective.
\newblock In: 2018 IEEE 25th Int. Conf. on Software Analysis, Evolution and
  Reengineering (SANER). IEEE; 2018. p. 26--37.

\bibitem{rattan2013software}
Rattan D, Bhatia R, Singh M.
\newblock Software clone detection: A systematic review.
\newblock Information and Software Technology. 2013;55(7):1165--1199.

\bibitem{tiarks2009assessment}
Tiarks R, Koschke R, Falke R.
\newblock An assessment of type-3 clones as detected by state-of-the-art tools.
\newblock In: Source Code Analysis and Manipulation, 2009. SCAM'09. Ninth IEEE
  Inter. Working Conf. on. IEEE; 2009. p. 67--76.

\bibitem{juergens2010achieving}
Juergens E, G{\"o}de N.
\newblock Achieving accurate clone detection results.
\newblock In: Proceedings of the 4th Inter. Workshop on Software Clones. ACM;
  2010. p. 1--8.

\bibitem{roy2009comparison}
Roy CK, Cordy JR, Koschke R.
\newblock Comparison and evaluation of code clone detection techniques and
  tools: A qualitative approach.
\newblock Science of computer programming. 2009;74(7):470--495.

\bibitem{svajlenko2014evaluating}
Svajlenko J, Roy CK.
\newblock Evaluating modern clone detection tools.
\newblock In: Software Maintenance and Evolution (ICSME), 2014 IEEE Int Conf
  on. IEEE; 2014. p. 321--330.

\bibitem{bellon2007comparison}
Bellon S, Koschke R, Antoniol G, Krinke J, Merlo E.
\newblock Comparison and evaluation of clone detection tools.
\newblock IEEE Trans on software eng. 2007;33(9).

\bibitem{thomsen2012clone}
Thomsen MJ, Henglein F.
\newblock Clone detection using rolling hashing, suffix trees and dagification:
  A case study.
\newblock In: Software Clones (IWSC), 2012 6th Intern. Workshop on. IEEE; 2012.
  p. 22--28.

\bibitem{guo2008detecting}
Guo J, Zou Y.
\newblock Detecting clones in business applications.
\newblock In: Reverse Engineering, 2008. WCRE'08. 15th Working Conference on.
  IEEE; 2008. p. 91--100.

\bibitem{ducasse1999language}
Ducasse S, Rieger M, Demeyer S.
\newblock A language independent approach for detecting duplicated code.
\newblock In: Software Maintenance, 1999.(ICSM'99) Proceedings. IEEE Inter Conf
  on. IEEE; 1999. p. 109--118.

\bibitem{baker1995finding}
Baker BS.
\newblock On finding duplication and near-duplication in large software
  systems.
\newblock In: Reverse Engineering, 1995., Proceedings of 2nd Working Conference
  on. IEEE; 1995. p. 86--95.

\bibitem{baxter1998clone}
Baxter ID, Yahin A, Moura L, Sant'Anna M, Bier L.
\newblock Clone detection using abstract syntax trees.
\newblock In: Software Maintenance, 1998. Proceedings., Int. Conf. on. IEEE;
  1998. p. 368--377.

\bibitem{krinke2001identifying}
Krinke J.
\newblock Identifying similar code with program dependence graphs.
\newblock In: Reverse Engineering, 2001. Proceedings. Eighth Working Conference
  on. IEEE; 2001. p. 301--309.

\bibitem{mayrand1996experiment}
Mayrand J, Leblanc C, Merlo E.
\newblock Experiment on the Automatic Detection of Function Clones in a
  Software System Using Metrics.
\newblock In: icsm. vol.~96; 1996. p. 244.

\bibitem{kamiya2002ccfinder}
Kamiya T, Kusumoto S, Inoue K.
\newblock CCFinder: a multilinguistic token-based code clone detection system
  for large scale source code.
\newblock IEEE Trans on Software Engineering. 2002;28(7):654--670.

\bibitem{sheneamer2015code}
Sheneamer A, Kalita J.
\newblock Code clone detection using coarse and fine-grained hybrid approaches.
\newblock In: 2015 IEEE seventh international conference on intelligent
  computing and information systems (ICICIS). IEEE; 2015. p. 472--480.

\bibitem{hummel2010index}
Hummel B, Juergens E, Heinemann L, Conradt M.
\newblock Index-based code clone detection: incremental, distributed, scalable.
\newblock In: 2010 IEEE International Conference on Software Maintenance. IEEE;
  2010. p. 1--9.

\bibitem{toomey2012ctcompare}
Toomey W.
\newblock Ctcompare: Code clone detection using hashed token sequences.
\newblock In: 2012 6th Inter Workshop on Software Clones (IWSC). IEEE; 2012. p.
  92--93.

\bibitem{uddin2011effectiveness}
Uddin MS, Roy CK, Schneider KA, Hindle A.
\newblock On the effectiveness of simhash for detecting near-miss clones in
  large scale software systems.
\newblock In: Reverse Eng. (WCRE), 2011 18th Working Conf on. IEEE; 2011. p.
  13--22.

\bibitem{white2016deep}
White M, Tufano M, Vendome C, Poshyvanyk D.
\newblock Deep learning code fragments for code clone detection.
\newblock In: Proceedings of the 31st IEEE/ACM Inter Conf on Automated Software
  Engineering. ACM; 2016. p. 87--98.

\bibitem{peng2018bayesian}
Peng M, Xie Q, Wang H, Zhang Y, Tian G.
\newblock Bayesian sparse topical coding.
\newblock IEEE Transactions on Knowledge and Data Engineering. 2018;.

\bibitem{grove2001framework}
Grove D, Chambers C.
\newblock A framework for call graph construction algorithms.
\newblock ACM Trans on Programming Languages and Systems (TOPLAS).
  2001;23(6):685--746.

\bibitem{mikhailov2016control}
Mikhailov A, Hmelnov A, Cherkashin E, Bychkov I.
\newblock Control flow graph visualization in compiled software engineering.
\newblock In: Information and Communication Technology, Electronics and
  Microelectronics (MIPRO), 2016 39th International Convention on. IEEE; 2016.
  p. 1313--1317.

\bibitem{sparks2007automated}
Sparks S, Embleton S, Cunningham R, Zou C.
\newblock Automated vulnerability analysis: Leveraging control flow for
  evolutionary input crafting.
\newblock In: Twenty-Third Annual Computer Security Applications Conference
  (ACSAC 2007). IEEE; 2007. p. 477--486.

\bibitem{ball1996efficient}
Ball T, Larus JR.
\newblock Efficient path profiling.
\newblock In: Proc. of the 29th annual ACM/IEEE inter symposium on
  Microarchitecture. IEEE Computer Society; 1996. p. 46--57.

\bibitem{wu1994static}
Wu Y, Larus JR.
\newblock Static branch frequency and program profile analysis.
\newblock In: Proceedings of the 27th annual international symposium on
  Microarchitecture. ACM; 1994. p. 1--11.

\bibitem{lim2014comparing}
Lim HI.
\newblock Comparing Control Flow Graphs of Binary Programs through Match
  Propagation.
\newblock In: 2014 IEEE 38th Annual Computer Software and Applications
  Conference. IEEE; 2014. p. 598--599.

\bibitem{bruschi2006detecting}
Bruschi D, Martignoni L, Monga M.
\newblock Detecting self-mutating malware using control-flow graph matching.
\newblock In: International Conference on Detection of Intrusions and Malware,
  and Vulnerability Assessment. Springer; 2006. p. 129--143.

\bibitem{nandi2016anomaly}
Nandi A, Mandal A, Atreja S, Dasgupta GB, Bhattacharya S.
\newblock Anomaly detection using program control flow graph mining from
  execution logs.
\newblock In: Proceedings of the 22nd ACM SIGKDD International Conference on
  Knowledge Discovery and Data Mining. ACM; 2016. p. 215--224.

\bibitem{allen1970control}
Allen FE.
\newblock Control flow analysis.
\newblock In: ACM Sigplan Notices. vol.~5. ACM; 1970. p. 1--19.

\bibitem{wang2014hashing}
Wang J, Shen HT, Song J, Ji J.
\newblock Hashing for similarity search: A survey.
\newblock arXiv preprint arXiv:14082927. 2014;.

\bibitem{wang2018survey}
Wang J, Zhang T, Sebe N, Shen HT, et~al.
\newblock A survey on learning to hash.
\newblock IEEE Trans on Pattern Analysis and Machine Intelligence.
  2018;40(4):769--790.



\bibitem{datar2004locality}
Datar M, Immorlica N, Indyk P, Mirrokni VS.
\newblock Locality-sensitive hashing scheme based on p-stable distributions.
\newblock In: Proc. of the twentieth annual symposium on Computational
  geometry. ACM; 2004. p. 253--262.

\newpage

\bibitem{charikar2002similarity}
Charikar MS.
\newblock Similarity estimation techniques from rounding algorithms.
\newblock In: Proceedings of the thiry-fourth annual ACM symposium on Theory of
  computing. ACM; 2002. p. 380--388.

\bibitem{raymond2002maximum}
Raymond JW, Willett P.
\newblock Maximum common subgraph isomorphism algorithms for the matching of
  chemical structures.
\newblock Journal of computer-aided molecular design. 2002;16(7):521--533.

\bibitem{manku2007detecting}
Manku GS, Jain A, Das~Sarma A.
\newblock Detecting near-duplicates for web crawling.
\newblock In: Proceedings of the 16th international conference on World Wide
  Web. ACM; 2007. p. 141--150.

\bibitem{jang2011bitshred}
Jang J, Brumley D, Venkataraman S.
\newblock Bitshred: feature hashing malware for scalable triage and semantic
  analysis.
\newblock In: Proceedings of the 18th ACM conference on Computer and
  communications security. ACM; 2011. p. 309--320.

\bibitem{alomari2013self}
Alomari FB, Menasc{\'e} DA.
\newblock Self-protecting and self-optimizing database systems: Implementation
  and experimental evaluation.
\newblock In: Proceedings of the 2013 ACM Cloud and Autonomic Computing
  Conference. ACM; 2013. p.~18.

\bibitem{alomari2012autonomic}
Alomari F, Menasce DA.
\newblock An autonomic framework for integrating security and quality of
  service support in databases.
\newblock In: 2012 IEEE Sixth Inter. Conf. on Software Security and
  Reliability. IEEE; 2012. p. 51--60.

\bibitem{peng2018mining}
Peng M, Zhu J, Wang H, Li X, Zhang Y, Zhang X, et~al.
\newblock Mining event-oriented topics in microblog stream with unsupervised
  multi-view hierarchical embedding.
\newblock ACM Transactions on Knowledge Discovery from Data (TKDD).
  2018;12(3):38.

\bibitem{peng2018block}
Peng M, Shi H, Xie Q, Zhang Y, Wang H, Li Z, et~al.
\newblock Block Bayesian Sparse Topical Coding.
\newblock In: 2018 IEEE 22nd International Conference on Computer Supported
  Cooperative Work in Design ((CSCWD)). IEEE; 2018. p. 271--276.

\end{thebibliography}

\end{document}